\documentclass[aps,pra,showpacs,showkeys,reprint,twocolumn,superscriptaddress,preprintnumbers,amsmath,amssymb]{revtex4-1}
\usepackage{graphicx}
\usepackage{dcolumn}
\usepackage{bm}
\usepackage{xspace} 
\usepackage{dsfont} 
\usepackage{psfrag}
\usepackage{bbm}
\usepackage{hyperref}
\usepackage{float}
\usepackage[caption=false]{subfig}
\usepackage[usenames]{color}
\usepackage{textcomp}

\begin{document}

\title{Analysis of photon triplet generation in pulsed cascaded parametric down-conversion sources}
\author{Stephan Krapick}
\email[corresponding author: ]{krapick@mail.uni-paderborn.de}
\author{Christine Silberhorn}
\affiliation{Applied Physics/Integrated Quantum Optics, Department of Physics, University of Paderborn, Warburger Str. 100, 33098 Paderborn, Germany}

\date{\today}

\begin{abstract}
We analyze the generation rates and preparation fidelities of photon triplet states in pulsed cascaded parametric down-conversion (PDC) under realistic experimental circumstances. As a model system, we assume a monolithically integrated device with negligible interface loss between the two consecutive PDC stages. We model the secondary down-conversion process in terms of a lossy channel and provide a detailed analysis of noise contributions. Taking variable pump powers into account, we estimate the impact of higher-order photon contributions and conversion processes on the achievable coincidence probabilities. At mean photon numbers of $\langle m\rangle\sim0.25$ photons per pulse behind the first PDC stage, we expect around $4.0$ genuine photon triplets per hour. Additionally, we discuss fundamental limitations of our model system as well as feasible improvements to the detectable photon triplet rate.
\end{abstract}


\maketitle

\begin{section}{Introduction}
\label{sec:01}
Multipartite entanglement has been studied \cite{Greenberger1989, Greenberger1990} and experimentally demonstrated (see, for example \cite{Bouwmeester1999}) in recent years. It has been shown that Greenberger-Horne-Zeilinger-states (GHZ) can be used in order to prove the nonlocality of quantum physics \cite{Pan2000} and, thus, the lack of hidden variables \cite{Einstein1935,Bell1964,Bell1966}.

In 2010, H\"ubel et al. \cite{Huebel2010} demonstrated the direct generation of photonic triplets, which are the fundamental states for multi-partite entangled state generation, using cascaded parametric down-conversion (PDC) sources. The same group also verified tripartite entanglement in the energy-time regime \cite{Shalm2013} as well as in the polarization degree of freedom \cite{Hamel2014}. Their experimental results have proven the feasibility of generating heralded Bell states with high visibility and fidelity, but without the need of post-selection. Likewise, the violation of Clauser-Horne-Shimony-Holt`s (CHSH) \cite{Clauser1969} inequality by over three standard deviations and the Mermin \cite{Mermin1990,Mermin1990a, Mermin1990b} inequality by more than ten standard deviations were demonstrated. With this type of cascaded PDC sources, hyper-entanglement of photon triplets or heralded GHZ states become achievable perspectives.

In these initial experiments continuous wave lasers were employed, while pulsed pump light would provide timing information beneficial for source synchronization and simplifying the analysis of higher photon number contributions.

Here, we study the photon triplet generation performance for the pulsed regime. In detail, we make general assumptions concerning the design parameters and measurement conditions, and we calculate not only the achievable rates of three-fold coincidences, but also show the significant influence of higher-order photon contributions on the measurement outcome. Furthermore, the analysis of the signal-to-noise behavior with real-world optical components and detectors is provided. Finally, we discuss optimization possibilities for the photon triplet source as well as for the measurement apparatus. We find that the measurable photon triplet rate can be improved applying the latest developments in detector technology (see, for example \cite{Lita2008,Marsili2013,Gerrits2012,Verma2015}) as well as pump lasers with high repetition rates.

The paper is structured as follows: In Section \ref{sec:01a}, we briefly describe the considered model system, which is followed by the estimation of primary PDC photon number output in Section \ref{sec:02}. Sections \ref{sec:03} and \ref{sec:03a} describe the secondary PDC process in terms of a lossy channel and the spectral splitting of the generated secondary photon pairs, respectively. In Section \ref{sec:04} we analyze the two- and three-fold coincidence probabilities, and we conclude in Section \ref{sec:05}
\end{section}

\begin{section}{Description of the model system}
\label{sec:01a}
Our theoretical and experimental considerations are inspired by the idea to generate photon triplets with cascaded PDC processes, as shown by H\"ubel et al. \cite{Huebel2010}. As a model system we consider the monolithic guided-wave device depicted in Fig. \ref{fig:01}. The input waveguide, assumed to be fabricated by titanium-diffusion for low loss \cite{Regener1985}, is periodically poled with two different grating periods in consecutive areas. In the first poled region, nondegenerate (primary) photon pairs at around $\lambda_\mathrm{s1}\approx790.5\,\mathrm{nm}$  (signal 1) and $\lambda_\mathrm{i1}\approx1625\,\mathrm{nm}$  (idler 1) are generated in a type-0 parametric down-conversion (PDC) process from a pulsed pump at $532\,\mathrm{nm}$. 

In order to separate the primary pair of photons in a spatio-spectral manner, we consider a passive wavelength division multiplexer (WDM) between the two poled regions. Idler 1 photons at the longer wavelength are transferred to an adjacent waveguide by evanescent field coupling with efficiency $\eta_\mathrm{coupler,i1}$, and they pass on-chip optical losses with efficiency $\eta_\mathrm{WG,i1}$. The WDM is wavelength selective in the telecom region for transverse magnetic (TM) polarization.
\begin{figure*}[ht]
\includegraphics[width=\linewidth]{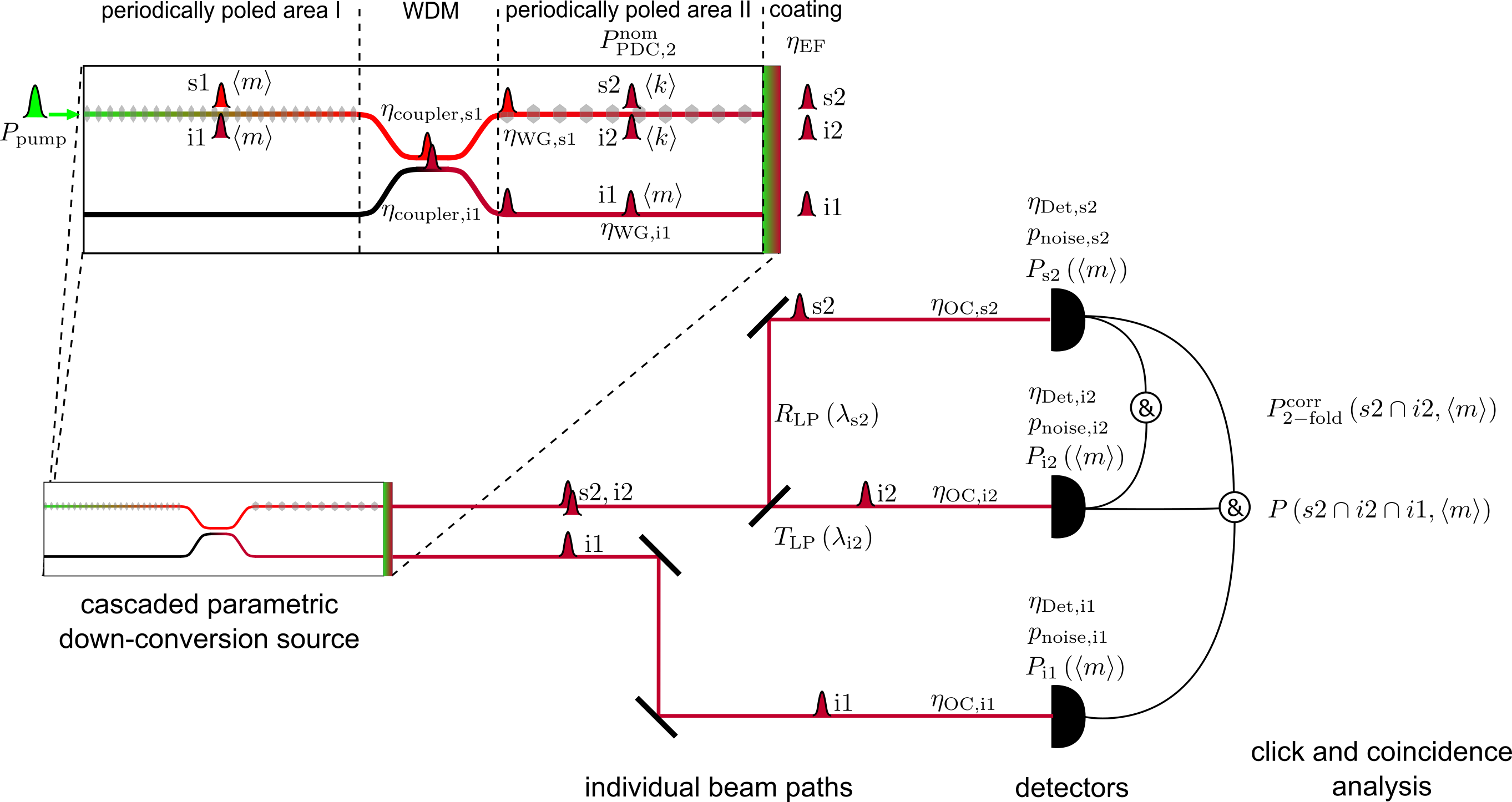}
\caption{\label{fig:01} Monolithic source and measurement schematic for cascaded parametric down-conversion. In the periodically poled area I, pulsed pump photons decay to nondegenerate photon pairs, s1 and i1, sharing the same mean photon number per pulse $\langle m\rangle$. The primary photon pair splits up spatio-spectrally at the integrated wavelength division multiplexer (WDM). Short-wavelength s1 photons remain in the original waveguide and decay to time-correlated secondary photon pairs, s2 and i2 (with common mean photon number per pulse $\langle k\rangle\ll\langle m\rangle$), in the periodically poled area II. Effectively, we convert one pump photon into three telecom photons, which are directed to individual detectors and undergo coincidence measurements, as described in detail in the text.}
\end{figure*}

The short-wavelength signal photons, generated in the fundamental spatial mode, remain in the pump channel with efficiency $\eta_\mathrm{coupler,s1}$ due to strong mode confinement, and they pass optical loss along the WDM and the straight waveguide with efficiency $\eta_\mathrm{WG,s1}$. 

The aforementioned components of the model system have been experimentally tested in \cite{Krapick2013,Krapick2014}, and we assume similar linear and nonlinear optical properties here.

In the secondary periodic poling, primary signal photons are predominantly present and decay with the nominal conversion probability $P^\mathrm{nom}_\mathrm{PDC,2}$ into secondary photon pairs (signal 2 and idler 2), which are distributed around the degeneracy wavelength of $\lambda_\mathrm{s,2}=\lambda_\mathrm{i,2}=1581\,\mathrm{nm}$. This has to be in accordance to the rules of energy and momentum conservation (phase-matching) for the secondary type-0 PDC process. We assume that detuning of the secondary down-conversion source allows for nondegenerate operation such that the overall cascaded decay of a green pump photon into three telecom photons is described by
\begin{eqnarray}
\label{eq:01}
532\,\mathrm{nm} & \rightarrow & 1625\,\mathrm{nm} + 1551\,\mathrm{nm} + 1611\,\mathrm{nm}\\
\mathrm{pump} & \rightarrow & \mathrm{idler\,1} + \mathrm{signal\,2} + \mathrm{idler\,2}.\nonumber
\end{eqnarray}

In the case of primary signal photons we expect loss figures of $\eta_\mathrm{WG,s1}\le0.3\,\mathrm{dB/cm}$ and $\eta_\mathrm{WG,i1}=\eta_\mathrm{WG,s2}=\eta_\mathrm{WG,i2}\le0.08\,\mathrm{dB/cm}$ for telecom photons. All generated telecom photons can exit the device through an end-face coating with efficiency $\eta_\mathrm{EF}\ge0.99$.

Although our model system is an integrated device, the following considerations apply also to bulk and hybrid cascaded parametric down-conversion sources, if they are pumped with a pulsed laser system. Figure \ref{fig:01} shows the principle measurement scheme as well as the variables used in this work.
\end{section}

\begin{section}{Estimation of the mean photon number in the primary PDC process}
\label{sec:02}

In this section we model the outcome of the primary parametric down-conversion process in terms of the click probability for idler 1 photons at binary detectors. We assume the signal and idler photon number per optical pulse to be Poisson-distributed, since type-0 PDC processes are typically multi-mode in the spectral domain \cite{Krapick2014}. 

The density vector element, which describes the probability that an $m$-photon state contributes to the average photon number $\langle m \rangle$ per signal or idler pulse, is given by
\begin{equation}
\label{eq:02}
\rho_m=\frac{e^{-\langle m\rangle}\cdot \langle m\rangle^{m}}{m!}.
\end{equation}

Depending on the pump power, i. e., the photon number density per pump pulse, we can deduce the mean photon number per output pulse fairly well by measuring idler 1 clicks events with a binary free-running detector. This is only valid under the precondition that the detector does not saturate due to dead-time effects. 

For the (lossless) case of an arbitrary $m$-photon input state impinging the idler 1 detector, we can model the click response of the detector. If no photons arrive ($m=0$), only noise counts can influence the zero-photon click probability, which reads
\begin{equation}
\label{eq:03}
P'_\mathrm{i1}\left(m=0\right)= p_\mathrm{noise,i1}.
\end{equation}
The noise count probability of a free-running detector, $p_\mathrm{noise,i1}$, can be deduced from the specified dark count rate $R_\mathrm{dc,i1}$, and from the count rate of additional blackbody photons, $R_\mathrm{bb}$. The latter becomes significant, if we heat up the cascaded PDC source to operation points around $170^\circ$C. We relate the sum of the two rates, $R_\mathrm{noise,i1}$, to appropriate timing units in order to get the noise probability $p_\mathrm{noise,i1}$. For pulsed systems, the pulse duration, in combination with the joint timing jitter of the involved devices, appears to be a reasonable timing choice.

Second, when we know that there is exactly one photon impinging to the detector ($m=1$), a click can occur (logical OR) due to a noise count or due to the primary idler photon itself with efficiency $\eta_\mathrm{Det,i1}$ or both, i. e., noise counts 'blind' the detector for the incoming photon. Therefore, the are treated as stochastically independent, and we get the click probability for the one-photon state:
\begin{equation}
\label{eq:04}
P'_\mathrm{i1}\left(m=1\right)= p_\mathrm{noise,i1}+\eta_\mathrm{Det,i1}- p_\mathrm{noise,i1}\cdot\eta_\mathrm{Det,i1}.
\end{equation}

As soon as there are (higher-order) $m$-photon states arriving at the detector, the latter will give only one click, since it is not photon-number resolving. The corresponding click probability is deduced as the counter event of the case, where the detector neither clicks due to noise events (that is $p=1- p_\mathrm{noise,i1}$), nor due to any of the impinging photons (i. e., $p'=(1-\eta_\mathrm{Det,i1})^m$). This corresponds to a logical NAND operation to the negated zero- and one-photon click probabilities, and we write
\begin{equation}
\label{eq:05}
P'_\mathrm{i1}\left(m\right)=1-(1- p_\mathrm{noise,i1})\cdot(1-\eta_\mathrm{Det,i1})^m.
\end{equation}
Since this formula is consistent with equations \ref{eq:03} and \ref{eq:04} for the cases $m=0$ and $m=1$, respectively, we generalize equation \ref{eq:05} for arbitrary $m$-photon states with $m\ge 0$.
 
We expect primary idler photons to undergo optical loss of $1-\eta_\mathrm{i1}$ on their path from the point of generation to the point of being detected. In detail, the transmittance $\eta_\mathrm{i1}$ can be calculated as the product of individual transmittances and efficiencies of lossy optical elements and the detector (logical AND), that is
\begin{equation}
\label{eq:05a}
\eta_\mathrm{i1}=\eta_\mathrm{int,1}\cdot\eta_\mathrm{EF}\cdot\eta_\mathrm{OC,i1}\cdot\eta_\mathrm{Det,i1}.
\end{equation}
Herein $\eta_\mathrm{int,1}=\eta_\mathrm{coupler,i1}\cdot\eta_\mathrm{WG,i1}$ denotes the intrinsic transmittance of our model device for idler 1 photons behind the first PDC stage. The factor $\eta_\mathrm{OC,i1}$ is the transmittance of supplementary optical components in the idler 1 measurement arm, and the detector efficiency is still labeled $\eta_\mathrm{Det,i1}$. With this in mind, we substitute equation \ref{eq:05} by the loss-dependent $m$-photon click probability
\begin{equation}
\label{eq:06}
P_\mathrm{i1}\left(m\right)=1-(1- p_\mathrm{noise,i1})(1-\eta_\mathrm{i1})^m.
\end{equation}

Third, we calculate the overall click probability $P_\mathrm{i2}\left(\langle m \rangle\right)$ for an ensemble measurement of optical pulses, carrying Poisson-distributed photons with mean photon number $\langle m \rangle$. We weight the individual $m$-photon contributions with their occupation probabilities $\rho_\mathrm{m}$ and get for the overall click probability
\begin{equation}
\label{eq:07}
P_\mathrm{i1}\left(\langle m \rangle\right)=\sum\limits^\infty_{m\ge0}\left[1-(1- p_\mathrm{noise,i1})(1-\eta_{i1})^m\right]\cdot\rho_m.
\end{equation}

We find that $P_\mathrm{i1}\left(\langle m \rangle\right)$ is equal to the noise count probability $p_\mathrm{noise,i1}$ for $\langle m \rangle\rightarrow 0$. Contrarily, for large mean photon numbers $\langle m \rangle\gg1$, $P_\mathrm{i1}\left(\langle m \rangle\right)$  asymptotically approaches unit click probability. Another consequence of equation \ref{eq:07} is its linearity with respect to $\langle m\rangle$ for low arm efficiencies $\eta_\mathrm{i1}\ll1$. This means that an $m$-photon state will yield a click about $m$-times more often than a one-photon state. Furthermore, there is a linear relation between $P_\mathrm{i1}\left(\langle m \rangle\right)$ and $\langle m \rangle$ for $\langle m \rangle\ll1$, where only zero- and one-photon components play a predominant role.

Next, we model the response of a binary detector for a real-world scenario. If $\eta_\mathrm{i1}$ is properly known, we can calculate the click rate of a free-running detector in the idler arm to be approximately
\begin{equation}
\label{eq:08}
R_\mathrm{i1}\approx R_\mathrm{noise,i1}+R_\mathrm{rep}\cdot P_\mathrm{i1}\left(\langle m \rangle\right),
\end{equation}
where $R_\mathrm{rep}$ is the pump laser repetition rate.

Comparing the measured click rates with the calculated ones, we can deduce the mean photon number of the first PDC process by solving equation \ref{eq:08} for $P_\mathrm{i1}\left(\langle m \rangle\right)$ and get
\begin{equation}
\label{eq:08a}
P_\mathrm{i1}\left(\langle m \rangle\right) = \frac{R_\mathrm{i1}-R_\mathrm{noise,i1}}{R_\mathrm{rep}}\approx \frac{R_\mathrm{i1}}{R_\mathrm{rep}}.
\end{equation}

For small arm efficiencies $\eta_\mathrm{i1}\ll1$, we can infer that the mean photon number per pulse is approximately
\begin{equation}
\label{eq:08b}
\langle m\rangle\approx \frac{R_\mathrm{i1}}{\eta_\mathrm{i1}\cdot R_\mathrm{rep}}.
\end{equation}

Note that we assumed in the measured click rates in equation \ref{eq:08a} to be at moderate levels, where neither dead-time effects of the detector nor its saturation become significant. The validity of the latter condition depends on the efficiency $\eta_\mathrm{i1}$. However, the overall count rate should exceed the noise count rate significantly, i. e., $R_\mathrm{i1}\gg R_\mathrm{noise,i1}$, for the approximations in equations \ref{eq:08a} and \ref{eq:08a} to hold.\\
\end{section}

\begin{section}{Secondary PDC process}
\label{sec:03}

Accessing the mean photon number of the first PDC process by measuring primary idler clicks also allows us to predict the outcome of the secondary PDC process, since the primary photons share temporal correlations and have the same mean photon number per pulse at the point of their generation. Thus, we are able to estimate the contributions of individual $m$-photon states in an ensemble of optical pulses. We can be sure that, at the exit of our primary PDC stage, the photon number distribution of the signal photons is not only Poisson-like, but also equal to the corresponding idler distribution, if we neglect different loss values inside the chip.

In the following, we model the transfer of signal 1 photons towards the secondary conversion stage as well as the second PDC process itself in terms of a lossy channel with subsequent deterministic pair generation. An incoming $m$-photon state behaves as if it is reflected (and lost), when no conversion occurs, but gets transmitted and duplicated, when a conversion takes place. \textit{After} conversion, we treat the generated pair photons as individual but time-correlated entities. 

We also assume that the spectral characteristics of secondary photon pairs can be tuned such that deterministic splitting is achieved. Experimentally, this could be realized by implementing a long-pass (LP) filter and by driving the secondary PDC at nondegenerate photon pair emission. 

The linear optical device efficiency, 
\begin{equation}
\label{eq:09a}
\eta_\mathrm{Dev,s1}=\eta_\mathrm{coupler,s1}\cdot\eta_\mathrm{WG,s1},
\end{equation}
describes the transmission of signal 1 photons through the integrated WDM towards the secondary PDC stage, while $\eta_\mathrm{coupler,s1}$ is the efficiency that the signal photon remains in the original coupler arm. The term $\eta_\mathrm{WG,s1}$ denotes the waveguide transmittance for the signal 1 photon from the average point of its generation to the average point of its decay to a secondary photon pair. 

The secondary PDC process takes place with the \textit{nominal} conversion efficiency per signal 1 photon $P^\mathrm{nom}_\mathrm{PDC,2}$, which includes the spectral overlap of the primary signal photon mode with the phase-matched pump mode of the secondary process. Note that the mutual spatio-spectral mode compatibility between signal 1 photons and the pump input of the second PDC process is important, but we restrict ourselves to include this precondition to $P^\mathrm{nom}_\mathrm{PDC,2}$. We multiply the probabilities/efficiencies accordingly, since they describe stochastically independent events, and we get for the \textit{internal} conversion probability per incoming signal 1 photon in the second PDC process
\begin{equation}
\label{eq:09b}
P_\mathrm{PDC,2}=\eta_\mathrm{Dev,s1}\cdot P^\mathrm{nom}_\mathrm{PDC,2},
\end{equation}
which acts similar to a ``loss factor'' on primary signal photons. Note that $\eta_\mathrm{Dev,s1}$ can be replaced by the transmittance of any interface as long as the considered model system is a bulk or hybrid cascaded parametric down-conversion source.

We write the resulting photon number transformation in a matrix representation:
\begin{equation}
\label{eq:10}
\vec{\rho}_{k}=\mathbf{L}_\mathrm{PDC,2}\cdot\vec{\rho}_m.
\end{equation}
The vector $\vec{\rho}_m$ comprises the incoming photon number occupation probabilities for different $m$-photon states according to equation \ref{eq:02}. The elements
\begin{equation}
\label{eq:11}
L_{km}=\dbinom{m}{k}P_\mathrm{PDC,2}^k\left(1-P_\mathrm{PDC,2}\right)^{m-k}, \; m\ge k
\end{equation}
contribute to the loss matrix $\mathbf{L}_\mathrm{PDC,2}$ and describe the probability that $k$ secondary photon pairs are generated, given an $m$-photon state in the signal 1 (and idler 1) mode is present behind the first PDC stage.

This leads to \textit{individual} photon number contributions $\rho_k$ of the time-correlated signal 2 and idler 2 photons. We write
\begin{equation}
\label{eq:12}
\rho_{k}=\sum\limits^{\infty}_{m\ge k\ge 0}\dbinom{m}{k}P_\mathrm{PDC,2}^k\left(1-P_\mathrm{PDC,2}\right)^{m-k}\cdot\rho_m.
\end{equation}

In the limit of unit internal conversion efficiency $P_\mathrm{PDC,2}=1$, we expect $\rho_k=\rho_m$ to hold true, whereas real-world conditions lead to drastically decreased mean photon numbers per pulse for secondary photons, i. e., $\langle k \rangle\ll \langle m \rangle$, since nominal conversion efficiencies are typically of the order of $10^{-5}\le P^\mathrm{nom}_\mathrm{PDC,2}\le10^{-10}$ pairs per input photon.\\
\end{section}

\begin{section}{Spectral splitting of secondary photons}
\label{sec:03a}

In order to perform coincidence measurements between secondary photons, we have to split them spatio-spectrally. A realistic case for quasi-deterministic separation of nondegenerate secondary photon pairs is given, if we use, for example, a long-pass filter with a steep cut-on edge as the splitting element. Additional band-pass filters could be inserted in both, the reflected and the transmitted beam path. This provides the reduction of noise events, which are related to the blackbody emission of a heated cascaded parametric down-conversion source.

We combine the wavelength dependent transmission at supplementary optical elements (filters, fiber-couplings et cetera) with the splitting behavior of the long-pass filter and with the detection efficiency in the two individual measurement arms:
\begin{equation}
\label{eq:13}
\eta_\mathrm{s2}\left(\lambda_\mathrm{s2}\right)=R_\mathrm{LP}\left(\lambda_\mathrm{s2}\right)\cdot\eta_\mathrm{OC,s2}\left(\lambda_\mathrm{s2}\right)\cdot\eta_\mathrm{Det,s2}\left(\lambda_\mathrm{s2}\right)\cdot\eta_\mathrm{EF}\left(\lambda_\mathrm{s2}\right)
\end{equation}
and
\begin{equation}
\label{eq:14}
\eta_\mathrm{i2}\left(\lambda_\mathrm{i2}\right)=T_\mathrm{LP}\left(\lambda_\mathrm{i2}\right)\cdot\eta_\mathrm{OC,i2}\left(\lambda_\mathrm{i2}\right)\cdot\eta_\mathrm{Det,i2}\left(\lambda_\mathrm{i2}\right)\cdot\eta_\mathrm{EF}\left(\lambda_\mathrm{i2}\right).
\end{equation}
Herein, the optical components provide transmittances $\eta_\mathrm{OC}\left(\lambda\right)$ in the respective beam path. Note that $\eta_\mathrm{EF}\left(\lambda_\mathrm{i2}\right)$ describes the wavelength-dependent transmittance of telecom photons at the device's end-facet. Individual detector efficiencies are labeled with $\eta_\mathrm{Det}\left(\lambda\right)$, the wavelength dependence of which must be taken seriously, if common InGaAs-based detectors are used.

The aforementioned considerations allow us to anticipate the click probabilities of binary detectors in the respective measurement arm. The derivation is similar to equation \ref{eq:07}, but we take the significant changes of the photon number distribution, caused by the second PDC process (see equation \ref{eq:12}), into account. Additionally, we pay attention to the spectral dependence of the transmittances in equations \ref{eq:13} and \ref{eq:14}. 

A click event in one of the free-running detectors is given as the counter event of having neither a dark count, nor signal/idler photons, respectively. For the click probabilities of single detection events in the respective secondary arm we write
\begin{widetext}
\begin{eqnarray}
\label{eq:17}
P_\mathrm{s2/i2}\left(\langle m \rangle\right) & = & \sum\limits^\infty_{k\ge0}\!\left\{1-\left(1-p_\mathrm{noise,s2/i2}\right)\left[1-\eta_\mathrm{s2/i2}\left(\lambda_\mathrm{s2/i2}\right)\right]^k\right\}\cdot\rho_k\\ 
\label{eq:17a}
& = & \sum\limits^\infty_{k\ge0}\!\left\{1-\left(1-p_\mathrm{noise,s2/i2}\right)\left[1-\eta_\mathrm{s2/i2}\left(\lambda_\mathrm{s2/i2}\right)\right]^k\right\}\times\sum\limits^\infty_{m\ge k}\!\dbinom{m}{k}P_\mathrm{PDC,2}^k\left(1-P_\mathrm{PDC,2}\right)^{m-k}\rho_m.
\end{eqnarray}
\end{widetext}
We consider $p_\mathrm{noise,s2/i2}$ to be the individual noise count probabilities of the signal 2 and idler 2 detectors. With the realistic assumption of $P^\mathrm{nom}_\mathrm{PDC,2}\approx P_\mathrm{PDC,2}\approx 10^{-7}$, we can conclude that all $\left(m\ge1\right)$-photon states of the primary PDC process will be converted mainly to $k=0$ and few $k=1$ photon contributions in the secondary PDC process.

In the following, we assume that the wavelength dependent arm efficiencies can be described by analytical expressions, being continuous for reasonable intervals around the expected signal/idler wavelengths, i. e., $\lambda^\mathrm{min}_\mathrm{s2/i2}\le\lambda_\mathrm{s2/i2}\le\lambda^\mathrm{max}_\mathrm{s2/i2}$. Proper intervals given by the filter bandwidths of, say, fiber-optic band-pass filters. Thus, we write for the overall arm efficiencies $\eta^\mathrm{tot}$, which are experimentally accessible:
\begin{equation}
\label{eq:18b}
\eta^\mathrm{tot}_\mathrm{s2}=\frac{\int\limits\limits^{\lambda^\mathrm{max}_\mathrm{s2}}_{\lambda^\mathrm{min}_\mathrm{s2}}\!\eta_\mathrm{s2}\left(\lambda_\mathrm{s2}\right)\mathrm{d}\lambda}{\lambda^\mathrm{max}_\mathrm{s2}-\lambda^\mathrm{min}_\mathrm{s2}}
\end{equation}
and
\begin{equation}
\label{eq:18c}
\eta^\mathrm{tot}_\mathrm{i2}=\frac{\int\limits\limits^{\lambda^\mathrm{max}_\mathrm{i2}}_{\lambda^\mathrm{min}_\mathrm{i2}}\!\eta_\mathrm{i2}\left(\lambda_\mathrm{i2}\right)\mathrm{d}\lambda}{\lambda^\mathrm{max}_\mathrm{i2}-\lambda^\mathrm{min}_\mathrm{i2}}
\end{equation}

We furthermore approximate $(1-P_\mathrm{PDC,2})^{m-k} \approx 1$ for reasonable internal conversion efficiencies. Additionally, we expect $p_\mathrm{noise,s2}\ll\eta^\mathrm{tot}_\mathrm{s2}$ and $p_\mathrm{noise,i2}\ll\eta^\mathrm{tot}_\mathrm{i2}$. 

With all these assumptions and preconditions, the evaluation of equation \ref{eq:17a} for both photon species yields the click probabilities for secondary PDC photons:
\begin{eqnarray}
\label{eq:19}
P_\mathrm{s2}(\langle m \rangle)&\approx& p_\mathrm{noise,s2}+\eta^\mathrm{tot}_\mathrm{s2}\sum\limits\limits^\infty_{m\ge1}m\cdot P_\mathrm{PDC,2}\cdot\rho_m\\
\label{eq:20}
&=& p_\mathrm{noise,s2}+\eta^\mathrm{tot}_\mathrm{s2}\cdot P_\mathrm{PDC,2}^\mathrm{gen}\\
\label{eq:20a}
&=&p_\mathrm{noise,s2}+\eta^\mathrm{tot}_\mathrm{s2}\cdot P_\mathrm{PDC,2}\cdot\langle m\rangle.
\end{eqnarray}
and
\begin{eqnarray}
\label{eq:21}
P_\mathrm{i2}(\langle m \rangle) &\approx& p_\mathrm{noise,i2}+\eta^\mathrm{tot}_\mathrm{i2}\sum\limits\limits^\infty_{m\ge1}m\cdot P_\mathrm{PDC,2}\cdot\rho_m\\
\label{eq:22}
&=& p_\mathrm{noise,i2}+\eta^\mathrm{tot}_\mathrm{i2}\cdot P_\mathrm{PDC,2}^\mathrm{gen}\\
\label{eq:22a}
&=&p_\mathrm{noise,i2}+\eta^\mathrm{tot}_\mathrm{i2}\cdot P_\mathrm{PDC,2}\cdot\langle m\rangle.
\end{eqnarray}

In equations \ref{eq:19} to \ref{eq:22a} we notice, that primarily generated higher-order $m$-photon states will each contribute to the secondary conversion process approximately $m$-times more often than one-photon states. Note that the term $P_\mathrm{PDC,2}^\mathrm{gen}$ labels the effective pair generation probability per optical pulse, whereas $P_\mathrm{PDC,2}$ is the pair generation probability per incoming pump photon.

We infer that individual click probabilities depend linearly not only on $p_\mathrm{noise,i2}$, $P_\mathrm{PDC,2}$, and on $\eta^\mathrm{tot}_\mathrm{s2/i2}$, but also on the mean photon number of the primary PDC output,$\langle m\rangle$, since 
\begin{equation}
\label{eq:22b}
\sum\limits\limits^\infty_{m\ge1}m\cdot\rho_m\equiv\langle m\rangle.
\end{equation}
This means that four parameters have an influence on the single-click probabilities $P(\langle m \rangle)$. While the noise probability has to be kept as low as possible for good signal-to-noise ratios (SNR), the efficiency $\eta^\mathrm{tot}_\mathrm{s2/i2}$ should be as high as possible. 

Increasing $P_\mathrm{PDC,2}$ to values higher than $10^{-6}$ is technologically hard in lithium niobate waveguide structures. This fact implies, for experiments with commercially available equipment, that the noise count probability is typically of the same order as $P_\mathrm{PDC,2}$, and we will not be able to identify photon triplets with good signal-to-noise ratios just by measuring the generation of secondarily generated signal or idler photons. Instead, we will make use of the temporal correlations of the triplet photons and perform coincidence measurements, as described in the following. This will lift the genuine triplets above the inevitable accidental background contributions.

Note that equations \ref{eq:19} and \ref{eq:21} reflect what we would also expect for direct pumping of the secondary PDC process with attenuated laser pulses instead of single photons from the primary PDC stage.

\end{section}

\begin{section}{Coincidence measurements}
\label{sec:04}

For the evaluation of two- and three-fold coincidences, we look at the possible outcome options: in practice, we must accept that clicks at free-running binary detectors will stem from noise contributions OR genuine PDC photons. 

We distinguish between eight cases as shown in Fig. \ref{fig:02}, where only case $A$ represents a genuine photon triplet. All other options must not be discarded, but have to be taken into account as uncorrelated noise contributions.
\begin{figure*}[ht]
\includegraphics[width=\linewidth]{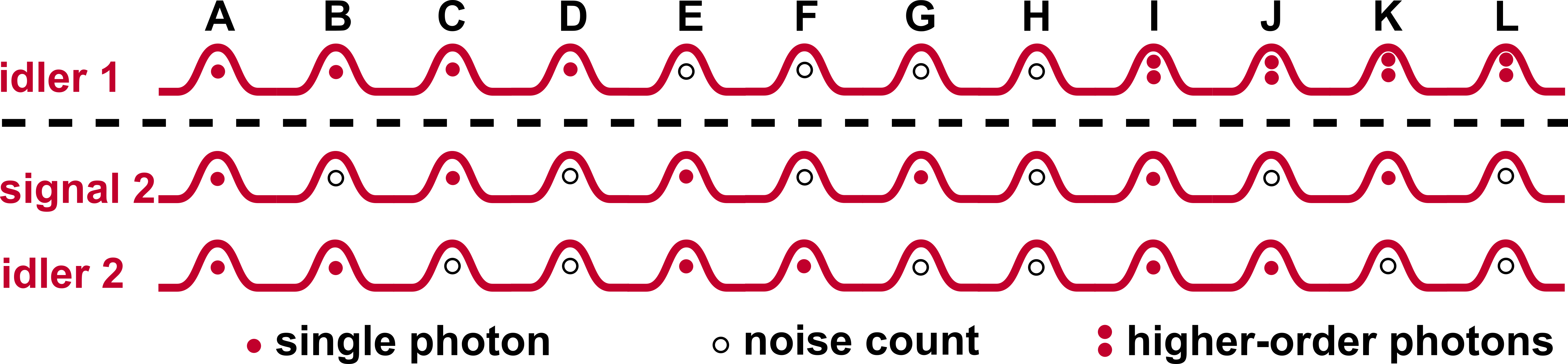}
\caption{\label{fig:02} Possible outcome of two- and three-fold coincidence measurements between the primary and secondary detection events. Only case A represents genuine photon triplets preparation, whereas all other options denote noise-related and accidental coincidences. The dashed line marks the spatial separation of events with high (top) and low (bottom) detection probabilities.}
\end{figure*}

\begin{subsection}{Two-fold coincidences of secondary PDC photons}
\label{subsec:41}

Parametric down-conversion processes in the pulsed regime generate time-correlated pairs of photons, which must not be treated as independent entities. In principle, the individual \textit{and} the coincidental detection probabilities both scale linearly with the pair generation probability $P_\mathrm{PDC}$. We write for the signal and idler probabilities
\begin{eqnarray}
\label{eq:25}
P_\mathrm{s} & = & \eta_\mathrm{s}\cdot P_\mathrm{PDC}\\
\label{eq:26}
P_\mathrm{i}\ & = & \eta_\mathrm{i}\cdot P_\mathrm{PDC},
\end{eqnarray}
and for the coincidence probability of correlated events we have
\begin{equation}
\label{eq:27}
P\left(\mathrm{s\cap i}\right)=P_\mathrm{i}\cdot P_\mathrm{s}\left(\mathrm{s|i}\right)=P_\mathrm{s}\cdot P_\mathrm{i}\left(\mathrm{i|s}\right)=\eta_\mathrm{s}\eta_\mathrm{i}\cdot P_\mathrm{PDC},
\end{equation}
where $\eta_\mathrm{s}$ and $\eta_\mathrm{i}$ represent the overall efficiencies in the respective detection arms. The detection of one photon, given that the detection of its twin already occurred, is expressed by the conditional probabilities
\begin{eqnarray}
\label{eq:28}
P_\mathrm{s}\left(\mathrm{s}|\mathrm{i}\right) & = & \frac{P\left(\mathrm{s\cap i}\right)}{P_\mathrm{i}}=\eta_\mathrm{K,s}=\eta_\mathrm{s}\\
\label{eq:29}
P_\mathrm{i}\left(\mathrm{i}|\mathrm{s}\right) & = & \frac{P\left(\mathrm{s\cap i}\right)}{P_\mathrm{s}}=\eta_\mathrm{K,i}=\eta_\mathrm{i}
\end{eqnarray}
The conditional detection probabilities (or Klyshko efficiencies $\eta_\mathrm{K}$ \cite{Klyshko1980}) are seemingly independent of the single photon pair generation probability, and they represent the respective overall path efficiencies. 

It has been shown experimentally that the coincidence probability scales super-linear for pump pulses, which contain several trillions of photons, while the single count probabilities scale almost linear with increasing pump powers \cite{Krapick2013}. This effect seems to be contradictory to the aforementioned considerations, but so far we implied the generation of twin photons only, while we neglected higher-order photon contributions appearing at higher pump powers.

In our case, we recognize the similarities of equations \ref{eq:20} and \ref{eq:22} and equations \ref{eq:25} and \ref{eq:26}, respectively. Hence, we infer the probability for the coincidental detection of the time-correlated secondary photons from equation \ref{eq:27}, and we write for the noiseless detection:
\begin{eqnarray}
\label{eq:29a}
P^\mathrm{corr}_\mathrm{2-fold}\left(\mathrm{s2\cap i2},\langle m\rangle\right)&=&\eta^\mathrm{tot}_\mathrm{s2}\eta^\mathrm{tot}_\mathrm{i2}\cdot P_\mathrm{PDC,2}^\mathrm{gen}\\
\label{eq:29b}
&=&\eta^\mathrm{tot}_\mathrm{s2}\eta^\mathrm{tot}_\mathrm{i2}\sum\limits\limits^\infty_{m\ge1}m \cdot P_\mathrm{PDC,2}\cdot\rho_m\\
\label{eq:29c}
&=&\eta^\mathrm{tot}_\mathrm{s2}\eta^\mathrm{tot}_\mathrm{i2}\cdot P_\mathrm{PDC,2}\cdot\langle m\rangle.
\end{eqnarray}
This means that the coincidence probability of secondary pairs scales linearly with the mean photon number of the primary process.

We estimate the noise-related accidental coincidences to be
\begin{eqnarray}
\label{eq:29d}
P^\mathrm{noise}_\mathrm{2-fold}\left(\mathrm{s2\cap i2},\langle m\rangle\right)&=&\eta^\mathrm{tot}_\mathrm{s2}\cdot P_\mathrm{PDC,2}\cdot\langle m\rangle\cdot p_\mathrm{noise,i2}\\\nonumber
& & +\eta^\mathrm{tot}_\mathrm{i2}\cdot P_\mathrm{PDC,2}\cdot\langle m\rangle\cdot p_\mathrm{noise,s2}\\\nonumber
& & + p_\mathrm{noise,s2}\cdot p_\mathrm{noise,i2},
\end{eqnarray}
and calculate the signal-to-noise ratio for secondary coincidence detection:
\begin{equation}
\label{eq:29e}
SNR_\mathrm{PDC,2}=\frac{P^\mathrm{corr}_\mathrm{2-fold}\left(\mathrm{s2\cap i2},\langle m\rangle\right)}{P^\mathrm{noise}_\mathrm{2-fold}\left(\mathrm{s2\cap i2},\langle m\rangle\right)}.
\end{equation}
This result indicates the feasibility to verify photon triplet generation by detecting coincidences only of secondary PDC photons, as long as the value of $SNR_\mathrm{PDC,2}$ is sufficiently high. However, for an imperfect triplet source, we have to consider the spatio-spectral splitting behind the first PDC stage to be nondeterministic. As long as implemented filters for secondary PDC wavelengths provide sufficient extinction at primary idler wavelengths, we can assume that the probability of detecting idler 1 photons in one or both of the secondary measurement arms is negligible and, thus, does not contribute to accidental coincidences of the secondary photon detection.

The situation is different and more critical in the case of higher-order mode combinations in the first PDC process of our model system. These can occur, if the $532\,\mathrm{nm}$ pump and/or the signal 1' photons are exited in, e. g., the $TM_{01}$ mode and propagating along the waveguide structure. Due to dispersion and different phase-matching conditions for higher-order PDC generation, we must assume that the corresponding idler 1' photons, although in the fundamental $TM_{00}$ mode, can have wavelengths similar to the secondary PDC photons, i. e., $\lambda_\mathrm{i1'}\approx\lambda_\mathrm{i1}$, which the on-chip coupler might not be optimized for. Given this case, the parasitic idler 1' photons could pass the WDM through the original arm (with transmittance $\eta^\mathrm{acc}_\mathrm{coupler,i1'}=1-\eta_\mathrm{coupler,i1'}(\lambda_\mathrm{i1'})$. They could also pass subsequent optical elements and band-pass filters towards the respective detector. Their contribution to secondary detection events is given by the probability, that at least one of the accidental idler 1' photon survives in the wrong measurement arm:
\begin{eqnarray}
\label{eq:29f}
P^\mathrm{acc}_\mathrm{i1'\rightarrow s2}&=&\eta^\mathrm{tot}_\mathrm{s2}\cdot\eta_\mathrm{WG,i1'}\cdot\eta_\mathrm{EF}\\\nonumber
&& \times\sum\limits^\infty_{m\ge0}\left[1-(1-\eta^\mathrm{acc}_\mathrm{coupler,i1'})^m\right]\cdot\rho_m'
\end{eqnarray}
and
\begin{eqnarray}
\label{eq:29g}
P^\mathrm{acc}_\mathrm{i1'\rightarrow i2}&=&\eta^\mathrm{tot}_\mathrm{i2}\cdot\eta_\mathrm{WG,i1'}\cdot\eta_\mathrm{EF}\\\nonumber
&& \times\sum\limits^\infty_{m\ge0}\left[1-(1-\eta^\mathrm{acc}_\mathrm{coupler,i1'})^m\right]\cdot\rho_m'.
\end{eqnarray}
For simplicity we assume that the internal conversion efficiency for the primary PDC process is equal for all spatial mode combinations. Thus, photon number occupation vector elements $\rho_m'=\rho_m$, identical to the ones of the fundamental PDC process, contribute to the formulas above. In practice, the conversion efficiencies strongly depend on the waveguide properties (i. e., the effective refractive indices of the involved modes) as well as on the possibility for quasi-selective excitation of different spatial pump modes.

We notice in equations \ref{eq:29f} and \ref{eq:29g} that the individual click probabilities for accidental idler 1' photons in one of the secondary measurement arms strongly depend not only on the properties of the integrated WDM structure, but also on the initial pump power and conversion efficiencies for primary higher-order PDC processes, both determining the individual photon number occupation densities $\rho_m'$. With the same arguments as in Section \ref{sec:02}, we deduce a linear increase of $P^\mathrm{acc}_\mathrm{i1\rightarrow s2}$ and $P^\mathrm{acc}_\mathrm{i1\rightarrow i2}$ with the pump power, i. e., the average photon number $\langle m'\rangle$, for small values of $\eta^\mathrm{acc}_\mathrm{coupler,i1}$.

Falsely directed idler 1' photons have an impact also on the coincidence click probability of the secondary process. We find that accidental coincidences occur with the joint probability
\begin{eqnarray}
\label{eq:29h}
P^\mathrm{acc,i1'}_\mathrm{coinc}\left(\mathrm{s2\cap i2},\langle m\rangle\right)&=&P^\mathrm{acc}_\mathrm{i1'\rightarrow s2}  \cdot P^\mathrm{acc}_\mathrm{i1'\rightarrow i2}\\\nonumber
&&+P^\mathrm{acc}_\mathrm{i1'\rightarrow s2}\cdot P_\mathrm{i2}(\langle m \rangle)\\\nonumber
&&+P^\mathrm{acc}_\mathrm{i1'\rightarrow i2}\cdot P_\mathrm{s2}(\langle m \rangle)\\\nonumber
&\approx&P^\mathrm{acc}_\mathrm{i1'\rightarrow s2}  \cdot P^\mathrm{acc}_\mathrm{i1'\rightarrow i2},
\end{eqnarray}
which includes coincidences of accidental idler 1' photons with noise contributions. For the approximation, we accounted for the typically low values of $P_\mathrm{PDC,2}\ll\eta^\mathrm{acc}_\mathrm{coupler,i1'}$. However, if we faithfully assume $\eta^\mathrm{acc}_\mathrm{coupler,i1'}\ll1$, we can approximate the coincidences-to accidental-ratios (CAR) for the secondary photon detection accordingly:
\begin{eqnarray}
\label{eq:29i}
CAR_\mathrm{PDC,2}&=&\frac{P^\mathrm{corr}_\mathrm{coinc}\left(\mathrm{s2\cap i2},\langle m\rangle\right)}{P^\mathrm{acc,i1'}_\mathrm{coinc}\left(\mathrm{s2\cap i2},\langle m\rangle\right)}\\\nonumber
&\approx& \frac{P_\mathrm{PDC,2}\cdot\langle m\rangle}{\eta_\mathrm{WG,i1'}^2\cdot\eta_\mathrm{EF}^2\cdot(\eta^\mathrm{acc}_\mathrm{coupler,i1'}\cdot\langle m'\rangle)^2}.
\end{eqnarray}
We notice that the coincidences-to-accidentals ratio decreases rapidly for higher pump powers, which denotes a fundamental limitation of an imperfect integrated device. Practical ways to overcome the decreasing fidelity is, consequently, to provide the suppression by higher-order PDC processes in the first stage by proper waveguide engineering. Linearly tapered waveguides, for example, can prevent higher-order pump modes at $532\,\mathrm{nm}$ to be guided. Likewise, tapered structures in between the integrated coupler and the secondary poled region can provide primary signal to be guided only in the fundamental spatial mode, whereas higher-order spatial modes at signal 1' wavelengths as well as telecom photons from the first PDC process scatter to the substrate due to the shallower refractive index profile of the waveguide behind tapering. These technological countermeasures can also reduce the necessity for strong and narrow-band filtering in the signal 2 and idler 2 measurement arms.
\end{subsection}

\begin{subsection}{Three-fold coincidences}
\label{subsec:42}

Although the coincidental detection of photon pairs generated in the second PDC process (provided that values of $SNR_\mathrm{PDC,2}$ and $CAR_\mathrm{PDC,2}$ are sufficiently high) can be seen as the sufficient condition to verify the generation of photon triplets, the full prove in terms of a necessary condition is given only, if we also detect the corresponding idler photon from the first PDC process in the dedicated measurement arm.

We rely on a measurement scheme, where all detector outcomes are fed into a multi-channel time-tagging unit, and we post-select the three-fold coincidences. In this configuration, we  benefit from having access to accidental events by analyzing coincidences between neighboring pulses \cite{Krapick2013}, which appear at multiple integers of the inverse repetition time of the pulsed pump laser.
\begin{figure*}[ht]
\includegraphics[width=\linewidth]{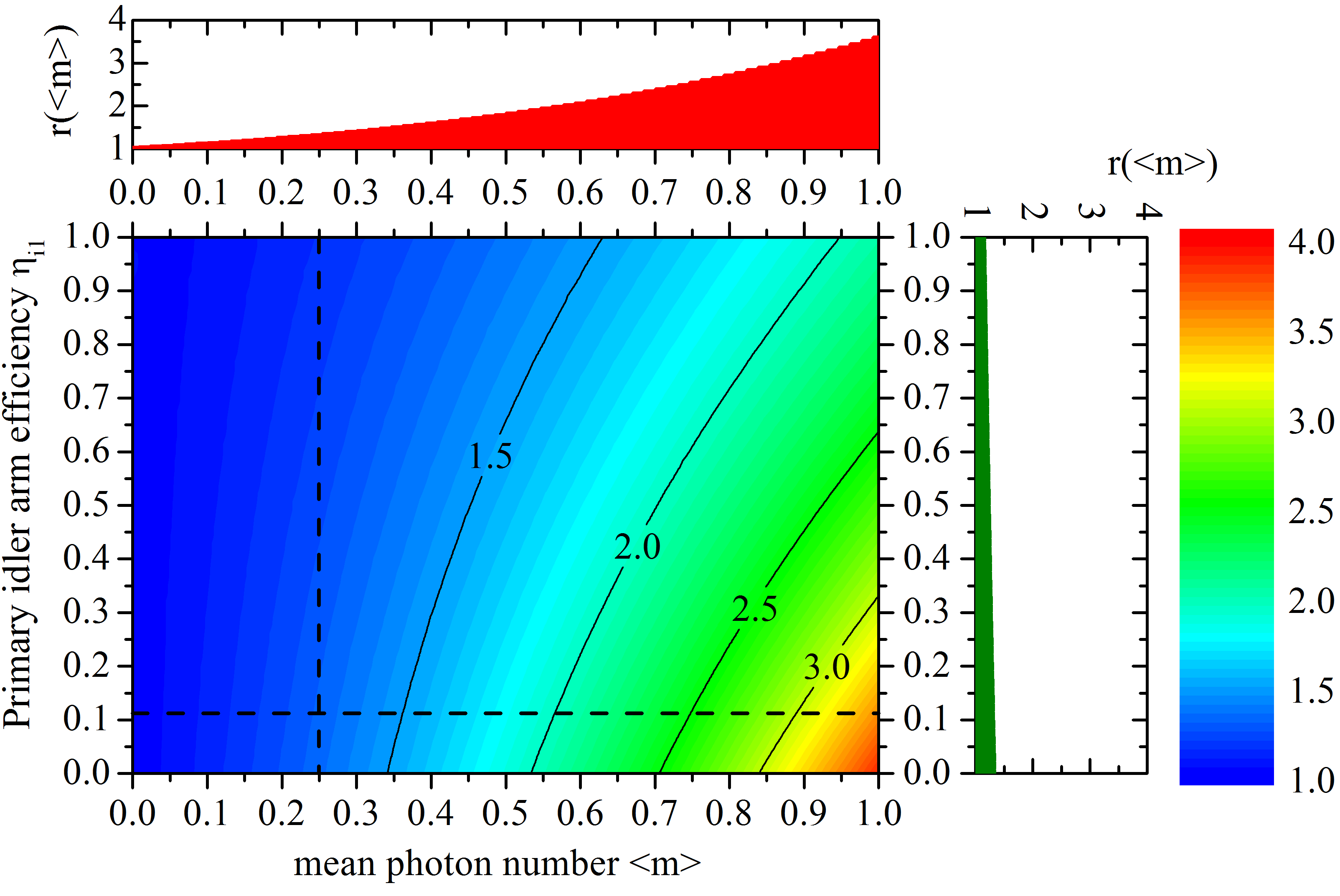}
\caption{\label{fig:03} Increase of the three-fold coincidence click probability dependent on the mean photon number and the idler arm efficiency of the primary PDC stage. The adjacent graphs show the individual dependencies for the dotted cross-cut lines, which represent reasonable values of $\eta_{i1}$ and $\langle m\rangle$. In general, we notice a nonlinear dependence of $r\left(\langle m\rangle\right)$ on both axis parameters, which represents a fundamental limitation on the generation and verification of genuine photon triplets in a pulsed system in conjunction with binary detectors.}
\end{figure*}

The three-fold coincidence probability of detecting time-correlated photons, which include genuine photon triplets, is given by
\begin{widetext}
\begin{eqnarray}
\label{eq:56b}
P\left(\mathrm{i2\cap s2\cap i1},\langle m\rangle\right)&=&\eta^\mathrm{tot}_\mathrm{i2}\cdot\eta^\mathrm{tot}_\mathrm{s2}\cdot\sum\limits^\infty_{m\ge1}m\cdot P_\mathrm{PDC,2}\cdot\left(1-P_\mathrm{PDC,2}\right)^{m-1}\cdot\left[1-\left(1-\eta_{i1}\right)^m\right]\cdot\rho_m\\
\label{eq:56c}
&\approx&\eta^\mathrm{tot}_\mathrm{i2}\cdot\eta^\mathrm{tot}_\mathrm{s2}\cdot P_\mathrm{PDC,2}\cdot\sum\limits^\infty_{m\ge1}m\cdot\left[1-\left(1-\eta_{i1}\right)^m\right]\cdot\rho_m.
\end{eqnarray}
The photon triplet detection probability is given by
\begin{eqnarray}
\label{eq:56d}
P\left(\mathrm{i2\cap s2\cap i1}, m=1\right)&\approx&\eta^\mathrm{tot}_\mathrm{i2}\cdot\eta^\mathrm{tot}_\mathrm{s2}\cdot P_\mathrm{PDC,2}\cdot\eta_{i1}\cdot\rho_1.
\end{eqnarray}
Thus, we can rewrite equation \ref{eq:56c} as
\begin{eqnarray}
\label{eq:56e}
P\left(\mathrm{i2\cap s2\cap i1},\langle m\rangle\right)&\approx&P\left(\mathrm{i2\cap s2\cap i1}, m=1\right)\cdot\left\{1+\sum\limits^\infty_{m\ge2}m\cdot\left[1-(1-\eta_{i1})^m\right]\cdot\frac{\rho_m}{\rho_1}\right\},
\end{eqnarray}
\end{widetext}
if we neglect noise contributions and assume, again, $(1-P_\mathrm{PDC,2})^{m-k} \approx 1$. We notice the influence of higher-order photons on the overall three-fold coincidence probability $P\left(\mathrm{i2\cap s2\cap i1},\langle m\rangle\right)$. Thus, we calculate the factor $r\left(\langle m\rangle\right)$ as the ratio of three-fold coincidences, that stem from $\langle m\rangle$-photon contributions, and $\left(m=1\right)$-photon events using equations \ref{eq:56d} and \ref{eq:56e}:
\begin{eqnarray}
r\left(\langle m\rangle\right) & = &\frac{P\left(\mathrm{i2\cap s2\cap i1}, \langle m\rangle\right)}{P\left(\mathrm{i2\cap s2\cap i1}, m=1\right)}\\\nonumber
\label{eq:57}
& \approx &1+\sum\limits^\infty_{m\ge2}m\cdot\left[1-(1-\eta_{i1})^m\right]\cdot\frac{\rho_m}{\rho_1}.
\end{eqnarray}
This indicates that higher-order photons in the first PDC stage have a nonlinear impact on the three-fold coincidences. We note that the ratio $\frac{\rho_m}{\rho_1}$ is linear only for $m\equiv2$. For $m\ge3$ we expect quasi-linear behavior only for low mean photon numbers $\langle m\rangle\ll1$, whereas the higher-order photon impact is super-linear otherwise.

The three-fold coincidence probability increase is depicted in Fig. \ref{fig:03}, where we plotted $r\left(\langle m\rangle\right)$ color-coded and dependent on the mean photon number $\langle m\rangle$ and the primary idler arm efficiency $\eta_{i1}$. We deduce from the adjacent graphs that, on the one hand, the idler 1 arm efficiency impacts $r\left(\langle m\rangle\right)$ quasi-linearly for $\langle m\rangle\approx0.25$. On the other hand, $r\left(\langle m\rangle\right)$ increases super-linearly with $\langle m\rangle$, which limits the scalability of a pulsed source fundamentally in terms of the injected pump power. Note that the inverse of $r\left(\langle m\rangle\right)$ is a measure for the ratio, with which genuine photon triplets contribute to the overall three-fold coincidence probability. This ratio decreases accordingly with increasing pump powers at the cascaded PDC source input.

Consequently, we treat three-fold coincidences, which involve higher-order photon contributions of the first PDC process, as accidentals. We have experimental access to the approximate value of $r\left(\langle m\rangle\right)$ by analyzing three-fold coincidences, which include neighboring pulses of the idler 1 detection at multiple integers of $R_\mathrm{rep}$. The probability of detecting three-fold coincidences in two consecutive pulses is equal to the probability of registering a three-fold coincidence between a secondary photon pair and a two-photon idler 1 pulse. Thus, for mean photon numbers $\langle m\rangle\le1$, where the higher-order photon states are dominated by two-photon contributions, we can approximate the accidentals in this manner. Note that this method is only possible in pulsed systems, whereas continuous-wave realizations would have to rely either directly on photon-number-resolving detectors, or on measurements of the second-order autocorrelation function of primary idler photons.

We calculate the achievable coincidences-to-accidentals ratio as
\begin{eqnarray}
\label{eq:57a}
CAR_\mathrm{3-fold} & = &\frac{P\left(\mathrm{i2\cap s2\cap i1}, m=1\right)}{\sum\limits^\infty_{m\ge2} P\left(\mathrm{i2\cap s2\cap i1}, m\right)}\\\nonumber
\label{eq:57b}
& = &\frac{1}{r\left(\langle m\rangle\right)-1},
\end{eqnarray}
which does not include noise-related contributions to the three-fold coincidences. Note that the coincidences-to-accidentals ratio has values $CAR_\mathrm{3-fold}<1$ for $r\left(\langle m\rangle\right)>2$. This means, in turn, that less than half of the three-fold coincidences stem from genuine photon triplets. We estimate $CAR_\mathrm{3-fold}\approx3.3$ for reasonably low pump powers, i. e., $\langle m\rangle=0.25$, and an idler 1 arm efficiency of $\eta_{i1}=0.117$.

The question, how many genuine photon triplet states we can expect to measure, is of interest now. From the approximation in equation \ref{eq:56d} and the efficiencies assumed for a realistic experimental setup (see Appendix), we derive for the expectable photon triplet rate:
\begin{eqnarray}
\label{eq:60}
R_\mathrm{triplet} & = & P\left(\mathrm{i2\cap s2\cap i1}, m=1\right)\cdot R_\mathrm{rep}\\
& = & \eta^\mathrm{tot}_\mathrm{i2}\cdot\eta^\mathrm{tot}_\mathrm{s2}\cdot P_\mathrm{PDC,2}\cdot\eta_\mathrm{i1}\cdot \rho_1\cdot R_\mathrm{rep}\nonumber\\
& \approx & \eta^\mathrm{tot}_\mathrm{i2}\cdot\eta^\mathrm{tot}_\mathrm{s2}\cdot P_\mathrm{PDC,2}\cdot R_\mathrm{i1}\nonumber\\
& =& 4.04 \;\text{h}^{-1},\nonumber
\end{eqnarray}
where a mean photon number of $\langle m\rangle=0.25$ per optical pulse in the first PDC process and a reasonable pump laser repetition rate of $R_\mathrm{rep}=10\,\mathrm{MHz}$ are included. From the result in equation \ref{eq:60} together with the individual measurement arm efficiencies, we derive a photon triplet generation rate inside the monolithic model system of $R_\mathrm{triplet}^\mathrm{gen}\sim1765\,\mathrm{h}^{-1}$. The theoretical generation rate is dependent on the repetition rate of the pump laser and the internal conversion efficiency of the secondary PDC process. Thus, it can be increased with appropriate technological improvements of the influencing parameters.

We assume picosecond pump pulses, and we expect the three-fold coincidences to be distributed over a temporal width of $\sim1\,\mathrm{ns}$, which includes the joint timing jitter of the detection apparatus. By analogy to the case of secondary photon pair coincidences, as discussed in Section \ref{subsec:41}, we derive the signal-to-noise ratio $SNR_\mathrm{3-fold}$ by relating the probability of generating three-fold coincidences by cascaded PDC, i. e., $P\left(\mathrm{i2\cap s2\cap i1},\langle m\rangle\right)$, to the probability of detecting noise-contributed three-fold coincidences. A valid approximation of $SNR_\mathrm{3-fold}$ includes three-fold coincidences, where the idler 1 detection events (PDC photons and noise contributions, see equation \ref{eq:07}) as well as the two-fold coincidences of the secondary photon detection are involved:
\begin{widetext}
\begin{eqnarray}
\label{eq:61}
P^\mathrm{noise}_\mathrm{3-fold}\left(\mathrm{i2\cap s2\cap i1},\langle m\rangle\right)&\approx&P^\mathrm{corr}_\mathrm{2-fold}\left(\mathrm{s2\cap i2},\langle m\rangle\right)\cdot p_\mathrm{noise,i1}+P^\mathrm{noise}_\mathrm{2-fold}\left(\mathrm{s2\cap i2},\langle m\rangle\right)\cdot P_\mathrm{i1}\left(\langle m\rangle\right)\\\nonumber
&=&\eta^\mathrm{tot}_\mathrm{s2}\eta^\mathrm{tot}_\mathrm{i2}\cdot P_\mathrm{PDC,2}\cdot\langle m\rangle\cdot p_\mathrm{noise,i1}+P^\mathrm{noise}_\mathrm{2-fold}\left(\mathrm{s2\cap i2},\langle m\rangle\right)\cdot\sum\limits^\infty_{m\ge0}\left[1-\left(1-\eta_{i1}\right)^m\right]\cdot\rho_m\\\nonumber
&\approx&P^\mathrm{noise}_\mathrm{2-fold}\left(\mathrm{s2\cap i2},\langle m\rangle\right)\cdot\sum\limits^\infty_{m\ge0}\left[1-\left(1-\eta_{i1}\right)^m\right]\cdot\rho_m.
\end{eqnarray}
\end{widetext}
The approximations in the formula are valid for the realistic case, where $SNR_\mathrm{PDC,2}\cdot p_\mathrm{noise,i1}\ll\eta_{i1}$ holds and $P^\mathrm{noise}_\mathrm{2-fold}\left(\mathrm{s2\cap i2},\langle m\rangle\right)\ll P^\mathrm{corr}_\mathrm{2-fold}\left(\mathrm{s2\cap i2},\langle m\rangle\right)\ll P_\mathrm{i1}\left(\langle m\rangle\right)$. For the upper bound of the signal-to-noise ratio of three-fold coincidences, we calculate
\begin{eqnarray}
\label{eq:62}
SNR_\mathrm{3-fold}&=&\frac{P\left(\mathrm{i2\cap s2\cap i1}, \langle m\rangle\right)}{P^\mathrm{noise}_\mathrm{3-fold}\left(\mathrm{i2\cap s2\cap i1},\langle m\rangle\right)}\\
&\approx&\frac{P\left(\mathrm{i2\cap s2\cap i1}, \langle m\rangle\right)}{P^\mathrm{noise}_\mathrm{2-fold}\left(\mathrm{s2\cap i2},\langle m\rangle\right)\cdot  P_\mathrm{i1}\left(\langle m\rangle\right)}\nonumber.
\end{eqnarray}
We see that $SNR\mathrm{3-fold}$ depends strongly on the probability of noise-related two-fold coincidences and, in turn, on the 
It is important to mention that this measure does not represent the signal-to-noise ratio of genuine photon triplets with respect to the noise-related background, but it rather depends on the overall three-fold coincidence probability. Thus, it is highly recommended to verify the generation of photon triplets by post-processing analysis of the signal-to-noise ratio \textit{and} the coincidences-to-accidentals ratio  in pulsed cascaded parametric down-conversion processes. 
\end{subsection}
\end{section}

\begin{section}{Conclusion and outlook}
\label{sec:05}
We presented the analysis of pulsed cascaded parametric down-conversion based on a monolithically integrated model system. For our real-world scenario, we derived the expectable two- and three-fold coincidence probabilities for the photon triplet detection process, where we included transmission and detection inefficiencies as well as noise contributions.

Taking reasonable experimental apparatus into account, we predicted photon triplet generation rates of $\sim4$ per hour, when the mean photon number of the primary PDC stage were $\langle m\rangle=0.25$ at laser repetition rates of $10\,\mathrm{MHz}$. The impact of uncorrelated noise contributions to the signal-to-noise ratio of three-fold coincidences has been shown, and we inferred that a careful analysis of higher-order photon contributions is as essential as the estimation of higher-order PDC processes in the first stage.
 
Significant improvements of the detectable photon triplet rate become feasible using highly efficient detectors with very low dark count rates with efficiencies of over $80\%$. Likewise, the influence of higher-order photon contributions on the primary idler detection rates will decrease using. Additionally, transition-edge detector systems operate in free-running mode with intrinsic photon number resolution. Thus, a convenient separation of genuine photon triplets from higher-order three-fold coincidences becomes an attractive option.

Higher repetition rates of the pump laser and the simultaneous reduction of the mean photon numbers per pump pulse are suitable options to furthermore reduce the impact of higher order photon contributions. Technological improvements to the individual PDC conversion efficiencies are also feasible using reverse-proton-exchanged waveguides \cite{Parameswaran2002}.\\
\end{section}

\begin{acknowledgments}
The authors would like to thank H. Herrmann and B. Brecht for helpful discussions. We also thank the Deutsche Forschungsgemeinschaft for funding this work within the Graduate Program ``Micro- and Nanostructures in Optoelectronics and Photonics'' (GRK 1464/2).
\end{acknowledgments}

\bibliography{Literature-Database}

\appendix*
\section{List of realistic conditions for our model sytem}

In this appendix, we give an overview on reasonable numbers for the properties of the model system, i. e., individual detection efficiencies, conversion efficiencies, transmittances, and detector noise characteristics for the calculations in the main text.
\begin{table*}
\caption{\label{tab:01} Summary of assumptions for our calculations}
\begin{ruledtabular}
\begin{tabular}{llll}
Optical element & characteristic & formula symbol & transmittance\\\hline
pump & pulse width & $\tau_\mathrm{pulse}$ & $4.4\cdot10^{-11}$ s\\
pump & repetition rate & $R_\mathrm{rep}$ & $10^{7}$ Hz\\
waveguide & endface transmittance & $\eta_\mathrm{EF}(\lambda\sim1590\,\mathrm{nm})$ & $0.995$\\\hline\hline
idler 1 & noise probability per ns & $p_\mathrm{noise,i1}$ & $7\cdot10^{-6}$\\
& detection efficiency & $\eta_\mathrm{Det,i1}(\lambda=1625\,\mathrm{nm})$ & $\sim0.45$\\
& coupler efficiency & $\eta_\mathrm{coupler,i1}(\lambda=1625\,\mathrm{nm})$ & $\sim0.94$\\
& waveguide transmittance & $\eta_\mathrm{WG,i1}(\lambda=1625\,\mathrm{nm})$ & $\sim0.92$\\
& optical components transmittance & $\eta_\mathrm{OC,i1}(\lambda=1625\,\mathrm{nm})$ & $\sim0.3$\\\hline
& \textbf{overall channel efficiency} & $\eta_\mathrm{i1}$ & $\sim0.117$\\\hline\hline
signal 1 & coupler efficiency & $\eta_\mathrm{coupler,s1}(\lambda=790.5\,\mathrm{nm})$ & $\sim0.995$\\
& waveguide transmittance & $\eta_\mathrm{WG,s1}(\lambda=790.5\,\mathrm{nm})$ & $\sim0.93$\\
& $2^{nd}$ PDC efficiency & $P^\mathrm{nom}_\mathrm{PDC,2}$ & $\sim2.7\cdot10^{-7}$\\\hline
& \textbf{internal process efficiency} & $P_\mathrm{PDC,2}$ & $\sim 2.52\cdot10^{-7}$\\\hline\hline
signal 2/idler 2 & integral transmittance of optical components & $\eta_\mathrm{OC,s2}$/$\eta_\mathrm{OC,i2}$ & $\sim0.411/\sim0.292$\\
& integral detector efficiency & $\eta_\mathrm{Det,s2}$/$\eta_\mathrm{Det,i2}$ & $\sim0.25/\sim0.65$\\
& noise probability per ns  & $p_\mathrm{noise,s2}$/$p_\mathrm{noise,i2}$ & $7.5\cdot10^{-6}$/$1.8\cdot10^{-5}$\\\hline
& \textbf{overall channel efficiency} & $\eta^\mathrm{tot}_\mathrm{s2}$/$\eta^\mathrm{tot}_\mathrm{i2}$ & $\sim0.103/\sim0.190$
\end{tabular}
\end{ruledtabular}
\end{table*}

\end{document}